\documentclass[aps,twocolumn,groupedaddress]{revtex4}

\usepackage{epsf,latexsym}
\usepackage{times}

\newcommand{\bra}[1]{\langle #1 |}
\newcommand{\ket}[1]{| #1 \rangle}

\newcommand{\pr}{Phys. Rev.}

\newcommand\SWAP{{\sf SWAP}}
\newcommand\CNOT{{\sf CNOT}}

\newcommand{\ra}{{\rightarrow}}
\newcommand{\up}{{\uparrow}}
\newcommand{\down}{{\downarrow}}

\newcommand{\be}{\begin{equation}}
\newcommand{\ee}{\end{equation}}
\newcommand{\ba}{\begin{eqnarray}}
\newcommand{\ea}{\end{eqnarray}}

\newcommand{\ignore}[1]{}

\def\CC{{\rm\kern.24em \vrule width.04em height1.46ex depth-.07ex
    \kern-.30em C}}
\def\P{{\rm I\kern-.25em P}}

\def\bbbone{{\mathchoice {\rm 1\mskip-4mu l} {\rm 1\mskip-4mu l}
{\rm 1\mskip-4.5mu l} {\rm 1\mskip-5mu l}}} 

\def\bbbc{{\mathchoice {\setbox0=\hbox{$\displaystyle\rm C$}\hbox{\hbox
to0pt{\kern0.4\wd0\vrule height0.9\ht0\hss}\box0}}
{\setbox0=\hbox{$\textstyle\rm C$}\hbox{\hbox
to0pt{\kern0.4\wd0\vrule height0.9\ht0\hss}\box0}}
{\setbox0=\hbox{$\scriptstyle\rm C$}\hbox{\hbox
to0pt{\kern0.4\wd0\vrule height0.9\ht0\hss}\box0}}
{\setbox0=\hbox{$\scriptscriptstyle\rm C$}\hbox{\hbox
to0pt{\kern0.4\wd0\vrule height0.9\ht0\hss}\box0}}}}

\def\bbbz{{\mathchoice {\hbox{$\sf\textstyle Z\kern-0.4em Z$}}
{\hbox{$\sf\textstyle Z\kern-0.4em Z$}}
{\hbox{$\sf\scriptstyle Z\kern-0.3em Z$}}
{\hbox{$\sf\scriptscriptstyle Z\kern-0.2em Z$}}}}

\newcommand{\putfig}[2]{$$\leavevmode\hbox{\epsfxsize=#2 cm
   \epsffile{#1.eps}}$$}
\newcommand{\insertfig}[2]{\leavevmode \vcenter{\hbox{\epsfxsize=#2 cm
   \epsffile{#1.eps}}}}

\begin{document}
\title{Spintronic devices as quantum networks}
\author{Radu Ionicioiu}
\affiliation{Quantum Information Group, Institute for Scientific Interchange (ISI), Viale Settimio Severo 65, I-10133 Torino, Italy}

\begin{abstract}
We explore spintronics from a quantum information (QI) perspective. We show that QI specific methods can be an effective tool in designing new devices. Using the formalism of quantum gates acting on spin and mode degrees of freedom, we provide a solution to a reverse engineering problem, namely how to design a device performing a given transformation between input and output. Among these, we describe an orientable Stern-Gerlach device and a new scheme to entangle two spins by transferring the entanglement from orbital to spin degrees of freedom. Finally, we propose a simple scheme to produce hyper-entangle electrons, i.e., particles entangled in both spin and mode degrees of freedom.
\end{abstract}

%\pacs{}
\maketitle

\section{Introduction}

Spintronics is enjoying an increased interest and considerable research effort has been directed into the field recently \cite{spintronics}. The pioneering Datta and Das spin transistor \cite{datta_das} has inspired a number of proposal for spin injection and detection devices. This development is significant from several perspectives. From a technological point of view, it provided new applications, including high density magnetic memories, new transistors and low-current switching devices. From a theoretical point of view, several new effects have been discovered and studied, like quantum spin Hall effect and spin Coulomb drag \cite{damico,weber}.

In the same time, the spin of a single particle has been considered as a qubit in several solid-state proposals for quantum computing \cite{kane,loss_ddv,burkard,barnes,aep_ri,taylor}. This parallel development of spintronics and quantum information processing (QIP) suggests that an interdisciplinary approach could be relevant and beneficial for both fields.

The purpose of this article is to explore the interface between spintronics and QIP. Specifically, we show how QIP methods and concepts (like quantum gates, encoding, entanglement swapping etc) can offer a new perspective and help to design new spintronic devices. As an example we apply these methods to a specific problem, namely entanglement. We explore several schemes for generating entanglement in spintronic devices and we examine entanglement between different degrees of freedom, like spin-spin, spin-orbit and entanglement transfer between these. We discuss {\em single-particle entanglement} (between spin and orbital degrees of freedom of a single particle), {\em two-particle entanglement} and we describe methods for spin-spin entanglement.

In a previous work \cite{pbs} we used the interplay between spin and mode degrees of freedom to design a mesoscopic polarizing beam splitter (PBS). The device, equivalent to a mesoscopic Stern-Gerlach, separates an unpolarized input current into two completely polarized output currents. The intuition behind the PBS design was to view it as a controlled-NOT gate between spin and mode degrees of freedom $\CNOT(\sigma, k)$. Here we develop a systematic framework based on this approach, namely we view spintronic devices as quantum networks acting on spin and orbital degrees of freedom. This enables us to build complex spintronic circuits out of a finite set of building blocks, in the same ways as complex quantum networks are build from elementary quantum gates. Thus the present framework can offer a solution to a reverse engineering problem: how to construct a spintronic device performing a given transformation between input and output.

A second motivation of this article comes from QIP. All previous proposals for solid state quantum computing with electrons (either static or mobile) considered only one degree of freedom, spin or charge (orbital). Charge-charge interactions between electrons are stronger than spin-spin ones. This makes easier to entangle two electrons using the Coulomb interaction, but it also makes the decoherence problem worse for charge qubits. Spin qubits have a longer decoherence time, but on the other hand entangling two spins is more difficult. In our approach, by taking into account both spin and orbital degrees of freedom, we can extend the previous architectures and improve their functionality. For example, we can start with two electrons entangled in modes (via Coulomb interaction) and then transfer the entanglement to spin degrees of freedom. Hence this approach can also offer inspiration for future QIP schemes.

\section{The electron as a qubit}

\subsection{Spin and orbital degrees of freedom}

In this section we discuss the formalism used in the article. The electron has been considered as a physical implementation for a qubit in several guises: spin or charge qubit, static or flying \cite{loss_ddv, quputer, bertoni, barnes, aep_ri}. Although some of the methods discussed here can also be applied to static qubits, our main focus will be on spintronics, hence on flying qubits.

We consider electrons to have two degrees of freedom, spin and orbital. In our picture the electron is injected in one of two parallel 1D quantum wires, labelled the 0- and the 1-mode, representing the orbital degree of freedom. We denote the state of a single electron by $\ket{\sigma; k}$, where $\sigma= \up, \down$ and $k=0, 1$ are, respectively, the spin and the orbital (mode) degrees of freedom. Hence a single electron can encode two qubits: a {\em spin qubit} $\ket{\sigma}$ and a {\em mode qubit} $\ket{k}$ \cite{mode_qudit}. We also assume the electron to be in the ground state of the quantum wire and we neglect higher energy levels (i.e., there is no interband coupling).

As we consider both orbital and spin degrees of freedom, the present approach can be viewed as a bridge/interface between previous proposals using flying qubits, either charge \cite{quputer,bertoni} or spin qubits \cite{aep_ri,barnes}.

Although extremely important from an experimental point of view, we will not discuss here decoherence and dissipation processes, as they are beyond the scope of this article. In the following we will simply assume that spin and charge transport are in the coherent regime.

\subsection{Quantum gates: universality} 

We now define the quantum gates we will use in the article. Let $\sigma_{x,y,z}$ be the Pauli matrices. Two important single qubit gates are the phase shift ${\sf P}(\varphi)$ and the Hadamard gate ${\sf H}$:
\begin{eqnarray}
{\sf P}(\varphi) &:=& \mbox{diag}\,(1,\, e^{i\varphi}) \cr
{\sf H} &:=& (\sigma_x+ \sigma_z)/\sqrt{2}
\end{eqnarray}
Equivalently, we can use single-qubit rotations around the $x$ and $z$ axes of the Bloch sphere:
\begin{eqnarray}
{\sf R_x}(\theta) &:=& e^{i\theta \sigma_x}= \cos \theta + i \sin \theta\  \sigma_x \cr
{\sf R_z}(\theta) &:=& e^{i\theta \sigma_z}= \cos \theta + i \sin \theta\ \sigma_z
\end{eqnarray}
The equivalence between the two sets can be established using the identities ${\sf P}(\varphi)= e^{i\varphi/2} {\sf R_z}(-\varphi/2)$, ${\sf H}= -i\, {\sf R_z}(\pi/4)\, {\sf R_x}(\pi/4)\, {\sf R_z}(\pi/4)$ and $e^{i\theta \sigma_x}= {\sf H} e^{i\theta \sigma_z} {\sf H}$. The single-qubit $\sf NOT$ gate which flips the qubit state is $\sigma_x= -i {\sf R_x}(\pi/2)$. All single qubit gates can act either on spin or on mode (orbital) degrees of freedom, and we will use a subscript to indicate this.

The most common two-qubit gates are the controlled-phase gate ${\sf C}(\varphi)$ and the $\CNOT$:
\begin{eqnarray}
{\sf C}(\varphi) &:=& \mbox{diag}\, (1,\,1,\,1,\,e^{i\varphi}) \cr
\CNOT &:=& \mbox{diag}\, (1,\,1,\,\sigma_x)
\end{eqnarray}
The relationship between these gates is $\CNOT= (\bbbone \otimes {\sf H})\cdot {\sf C}(\pi)\cdot (\bbbone \otimes {\sf H})$. In the computational basis, the action of the $\CNOT(i,j)$ gate between qubit $i$ (control) and $j$ (target) is given by the mapping $\ket{x,y}_{ij} \mapsto \ket{x, y \oplus x}_{ij}$, i.e., it flips the value of the target qubit iff the control qubit is 1.

In order to perform an arbitrary unitary operation on a $n$-qubit register (i.e., achieve universality), we need to implement a universal set of quantum gates. There are several such universal sets, like $\{ \sf H, P(\varphi), C(\pi) \}$, $\{ \sf H, P(\varphi), \CNOT\}$ or $\{ \sf H, C(\varphi) \}$. 

The first step is to find the spintronic implementation of these gates which will be used next as building blocks for more complex networks, like $\SWAP(\sigma; k)$, Bell state analyzer, entanglement generation/swapping etc.

Once we have an implementation of a universal set of gates, we can construct any unitary transformation on an arbitrary number of qubits. This gives the solution to the reverse engineering problem mentioned before: decompose the given transfer function between input and output into a set of elementary quantum gates.

\subsection{Single-qubit gates}

\subsubsection{Spin qubits}

One way of implementing an $SU(2)$ spin rotation is by using an external magnetic field. However, as this is not very practical in implementations, we use an alternative method by employing only {\em static} electric fields. Arbitrary single-qubit rotations can be realized using Rashba-active regions \cite{rashba, datta_das, aep_ri}. A particle with magnetic moment moving in a region with {\em static} electric field $\bf E$ will see a magnetic field ${\bf B= v \times E}/c^2$ which couple to its spin. The spin-orbit Hamiltonian is
\be
H_{so}= \alpha\, \vec \sigma \cdot (\bf v \times E)
\label{Hso}
\ee
where $\vec \sigma$ is the vector of Pauli matrices and $\alpha= g_m e \hbar/(4m c^2)$ ($g_m$ is the gyromagnetic factor). 

Suppose the particle moves in a 1D quantum wire and a static electric field $E_y$ is applied perpendicular to the confinement $xz$-plane. This will induce a spin rotation which depends on the direction of propagation: if the electron moves parallel to the $x$- ($z$-) axis, $H_{so}$ will produce a spin rotation $\sf R_z(\theta)_\sigma$ ($\sf R_x(\theta)_\sigma$, respectively). Alternatively, if the electron moves only in a straight line along $Oy$, applying static electric fields along $E_z$ and $E_x$ results in rotations in the spin space ${\sf R}_{\sf x}(\theta)_\sigma \otimes \bbbone_k$ and ${\sf R}_{\sf z}(\theta)_\sigma \otimes \bbbone_k$, around the $x$ and respectively, $z$ axis; the subscripts $\sigma, k$ indicate the subspace/qubit on which the gate acts, e.g., spin or mode.

A rotation in the spin space is implemented by having Rashba active regions on {\em both} 0- and 1-modes (e.g., using a single pair of top/bottom gates acting on both quantum wires) like in the next figure:
\putfig{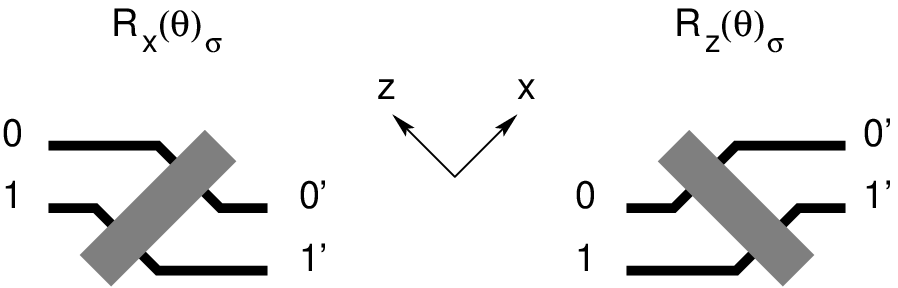}{6}
In the following quantum network diagrams we use red/dashed (blue/solid) lines for the spin (mode) qubits, respectively. In spintronic diagrams, we use black thick lines for the quantum wires representing the 0- and 1-modes.

It is important to note that the spin-orbit Hamiltonian (\ref{Hso}) is non-dispersive, hence the rotation angle $\theta$ does not depend on the velocity of the propagating electron. This is easy to understand: the phase $\theta \sim \int \vec \sigma 
\cdot ({\bf v \times E}) dt = \int {\bf dl}.(\vec \sigma \times {\bf E})$ is proportional to the length of the Rashba region and to the applied electric field $E$, and does not depend on $\bf v$. Another way to see this is that the Hamiltonian (\ref{Hso}) produces an Aharonov-Casher phase, and hence $\theta$ is a topologic phase \cite{top_phases}. Thus the rotations produced by $H_{so}$ have an build-in resilience to some type of errors.

\subsubsection{Mode qubits}

For mode qubits, all $SU(2)$ gates can be performed using beam-splitters and Aharonov-Bohm phases. A magnetic flux $\Phi$ confined between mode 0 and mode 1 generates an Aharonov-Bohm phase $\varphi= e\Phi/\hbar c$ \cite{ab}:
\[ \bbbone_\sigma\otimes {\sf P}(\varphi)_k:\ \ \ \ \ \cases{\ket{\sigma; 0} \ra\ \ket{\sigma; 0}\cr \ket{\sigma; 1} \ra\ e^{i\varphi}\ket{\sigma; 1}} \]
(note that the spin is unaffected if the flux is confined between the two quantum wires).

Let $a^\dag_{i,\sigma} (a_{i,\sigma})$, $i=0,1$ be the creation (annihilation) operator of an electron in mode $i$ and spin $\sigma$. A (spin insensitive) beam-splitter is described by the tunnelling (hopping) Hamiltonian
\be
H_h= \tau(t)a^\dag_{0,\sigma} a_{1,\sigma}+ \mbox{H.c.}
\ee
where $\tau(t)$ is the tunnelling rate. The unitary evolution given by $H_h$ acting on the system during $t \in [0,\, T]$ is a single qubit rotation around the $x$-axis in the $k$-space, with an angle $\theta= -\int_0^T \tau(t) dt /\hbar$:
\[ \bbbone_\sigma\otimes {\sf R}_{\sf x} (\theta)_k: \ \ \ \ \ \ \ket{\sigma; k} \ra\ \cos\theta\, \ket{\sigma; k} + i\sin\theta\, \ket{\sigma; 1-k} \]
with $k=0,1$.

A schematic of the spintronic circuits implementing these gates are in the next figure:
\putfig{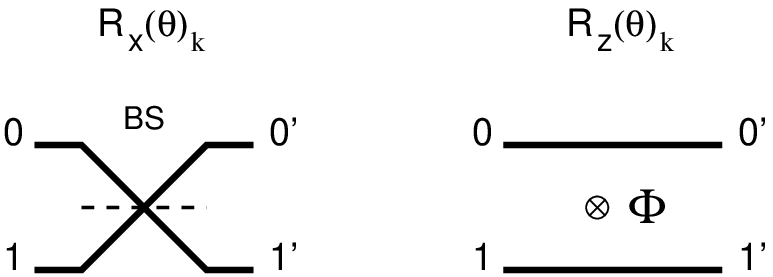}{6}

\begin{figure}
\putfig{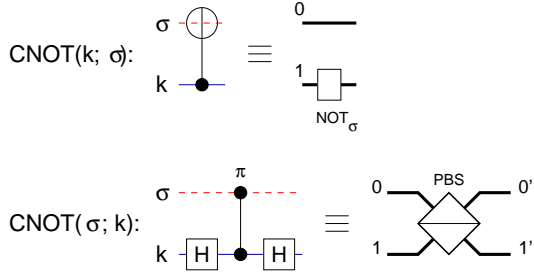}{7}
\caption{Quantum networks and equivalent spintronic circuits for the $\CNOT$ gate between spin (red, dashed line) and orbital (blue, solid line) degrees of freedom; 0 and 1 are the modes $k$. Top: $\CNOT(k; \sigma)$ gate: a spin-flip gate ${\sf NOT}_\sigma$ applied only to the mode $k=1$. Bottom: $\CNOT(\sigma; k)$ is equivalent to a polarizing beam splitter PBS.}
\label{cnot}
\end{figure}

\subsection{Two-qubit gates}

We now turn to the two-qubit gates. In this case, as we have different choices for the two qubits (either mode or spin), this will result in very different quantum networks.

A $\CNOT(k, \sigma)$ gate with the control on the mode qubit $k$ is simply a $\sf NOT_\sigma$ applied only on the 1-mode: if $k=1$, perform $\sf NOT_\sigma$; if $k=0$, do nothing (see Fig.\ref{cnot}, top). More generally, any mode-controlled $U_\sigma$ gate is just a $U_\sigma$ applied only to the 1-mode.

The ``reversed-controlled" gate $\CNOT(\sigma, k)$ has a different spintronic circuit (Fig.\ref{cnot}, bottom). We can understand its action better if we realize that a $\CNOT(\sigma, k)$ gate is equivalent to a polarizing beam splitter (PBS) \cite{pbs}. A spin $\up$ ($\down$) incident in mode $k$ will be transmitted to the same mode (reflected in the opposite mode):
\be
\CNOT(\sigma, k):\  \cases{ \ket{\up;\, k} \ra\ \ket{\up;\, k} \cr 
\ket{\down; k} \ra\ \ket{\down; 1-k}}
\ee

The spintronic circuit implementing the PBS is a Mach-Zehnder interferometer with an Aharonov-Bohm flux confined to its center and a Rashba active region only on the 1-mode enacting ${\sf R}_{\sf z}(\pi/2)_\sigma$. Due to the Rashba gate, spin up and spin down electrons will pick up a phase difference in the interferometer and therefore they will exit (with unit probability) through different output modes (see \cite{pbs} for more details about implementation).

\begin{figure}
\putfig{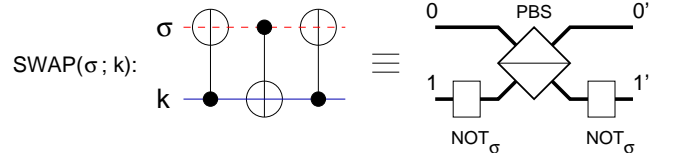}{8.5}
\caption{Schematics of a $\SWAP(\sigma, k)$ gate swapping the states of the spin and mode qubit. The gate is made of a polarizing beam splitter (PBS) sandwiched between two $\sf NOT_\sigma$ acting only on the 1-mode.}
\label{swap}
\end{figure}

Another useful two-qubit gate is the $\SWAP(i,j)$, whose action in the computational basis is $\ket{x,y}_{ij} \mapsto \ket{y,x}_{ij}$. It can be obtained from three $\CNOT$s in two equivalent ways: $\CNOT(i,j) \CNOT(j,i) \CNOT(i,j)$ or $\CNOT(j,i) \CNOT(i,j) \CNOT(j,i)$. For the $\SWAP(\sigma, k)$ gate, these two expressions lead to completely different spintronic circuits (again, due to the asymmetry between $\sigma$ and $k$). Since the mesoscopic PBS has four quantum gates (two beam-splitters, and two phases, Rashba and Aharonov-Bohm), the first version of the $\SWAP(\sigma, k)$ is made up of nine gates (two PBS and a $\sf NOT_\sigma$ gate), whereas the second version has only six gates (two $\sf NOT_\sigma$ gates and a PBS), see Fig.~\ref{swap}.

With the gates described above one can generate any unitary transformation on the spin and mode qubits of a single electron. In order to couple two different electrons $\ket{\sigma_1; k_1}$ and $\ket{\sigma_2; k_2}$, we need a two-qubit gate acting between a degree of freedom of the first electron and a degree of freedom of the second one.

Conceptually, the simplest such gate is a mode-mode coupling between the mode qubits of both electrons. This gate, known as {\em Coulomb coupler}, uses the Coulomb interaction between two electrons to perform a controlled phase gate ${\sf C}(\varphi)_{k_1 k_2}$ on the mode qubits \cite{quputer,bertoni,bell_test}, as in the next figure:
\putfig{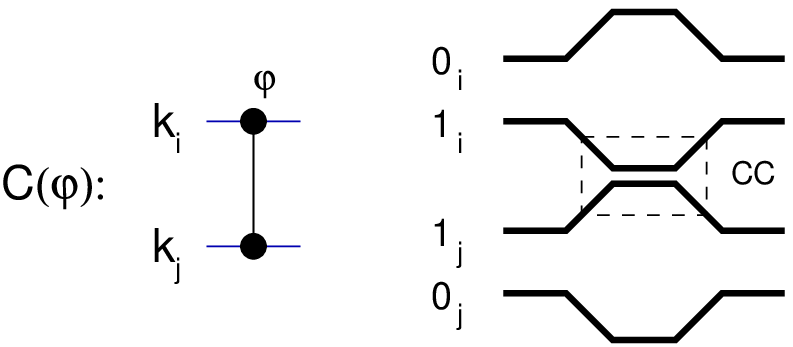}{6}
After the gate, only the $\ket{11}$ state picks up a phase $e^{i\varphi}$ due to the Coulomb interaction, since both electrons are in mode 1.

From the above discussion we see that single electron gates (i.e., those between $\sigma$ and $k$ of the same electron) are ``easy'', since they imply only single particle interactions. The electron propagates through predefined areas with different potential profiles (beam-splitters, phase shifters or Rashba active regions) and hence no synchronization is required. On the other hand, two-electrons gates like the Coulomb coupler are ``difficult'', since they are based on two-particle interaction and in this case synchronization is essential: the two electrons should reach the gate region simultaneously. One way to achieve this is to use a surface acoustic wave (SAW) \cite{barnes}. The electrons are trapped in the SAW minima and are pushed together through the whole circuit. In this case a SAW has a double role: it acts as a pump, pushing the electrons through the wires and also as a synchronizer, ensuring that different electrons arrive at the interaction region simultaneously.

This concludes the list of spintronic circuits implementing a universal set of gates, so now we can focus on building more complex networks. The next problem we discuss is how to produce different types of entanglement in these devices.

\section{Entanglement generation}

Starting with two unentangled qubits in a product state $\ket{x, y}$, $x,y= 0,1$, we can entangle them by applying the transformation $\CNOT(1,2) ({\sf H}_1\otimes \bbbone_2)$: first put the control qubit (say qubit 1) in a superposition using a Hadamard, ${\sf H}_1$, and then apply a controlled-{\sf NOT} gate. This transformation maps the computational basis $\{ \ket{x,y},\ x,y=0,1 \}$ to the Bell basis $\{ \ket{\Phi^\pm}, \ket{\Psi^\pm} \}$:
\begin{eqnarray}
\ket{\Phi^\pm}= (\ket{00} \pm \ket{11})/\sqrt{2} \cr
\ket{\Psi^\pm}= (\ket{01} \pm \ket{10})/\sqrt{2}
\end{eqnarray}
A Bell state analyzer performs a projective measurement on the Bell basis of an unknown two-qubit state $\ket{\psi}$. This is done by reversing the previous quantum network: we apply to the state $\ket{\psi}$ the transformation $({\sf H}_1\otimes \bbbone_2) \CNOT(1,2)$, followed by a measurement of both qubits in the computational basis $\{ \ket{x, y} \}$.

We will now apply this scheme to entangle two degrees of freedom of a single electron or of different electrons. 

\subsection{Spin-mode entanglement for a single electron}

It turns out that generating single particle $\sigma$-$k$ entanglement is relatively easy: we need only a beam-splitter followed by a $\sf NOT_\sigma$ gate on the 1-mode. A 50/50 beam splitter performs a ${\sf R_x}(\pi/4)_k= 2^{-1/2}\pmatrix{1& i \cr i& 1}$ rotation on the mode qubit; a spin flip gate $\sf NOT_\sigma$ acting only the 1-mode is equivalent to a $\CNOT(k, \sigma)$ gate. Therefore, the circuit entangling $k$ and $\sigma$ is:
\putfig{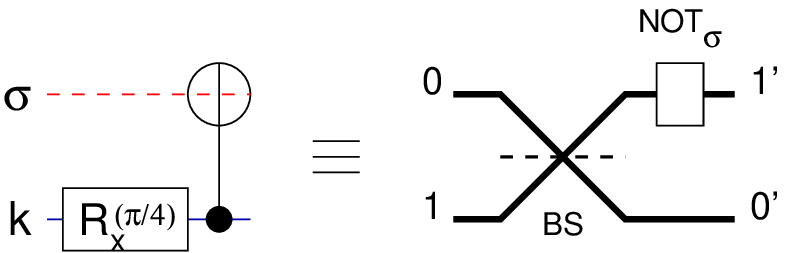}{6}
If we start with a single electron in the initial state $\ket{\up; 0}$, this device entangles the spin and orbital degrees of freedom by mapping the initial state to $\ket{\up; 0}\ra\ 2^{-1/2}(\ket{\up; 0}+ i \ket{\down; 1})=: \ket{\tilde\Psi^+}_{\sigma k}$ which is a maximally entangled state in spin-mode degrees of freedom.

\subsection{Two-particle mode-mode entanglement}

The Coulomb coupler discuss above is the key ingredient to produce mode-mode entangled electrons.

Starting with two electrons in $\ket{00}_{k_1 k_2}$ state (i.e., both electrons injected in the 0-rail) we first apply 50/50 beam-splitters on both electrons, then the Coulomb coupler and finally another beam-splitter only to the second electron. The resulting state is a mode-entangled Bell state $i \ket{\Psi^+}_{k_1 k_2}$ \cite{quputer,bertoni}. This architecture can be used to test the Bell inequality with ballistic electrons injected in quantum wires \cite{bell_test}.

Once we have a pair of mode-entangled electrons, we can transfer the mode-entanglement to spin-entanglement using known QIP protocols.

\subsection{Two-particle spin-spin entanglement}

Entangling the spins of two particles is challenging and several proposals have been put forward recently. In the following we show two schemes for spin entanglement.

\noindent{\bf (a) Entanglement swapping.} Suppose we have two pairs of entangled qubits, $(A_1, B_1)$ and $(A_2, B_2)$. If we perform a projective measurement on the Bell basis of qubits $B_1$ and $B_2$, the remaining two qubits, $A_1$ and $A_2$ become now entangled, a protocol known as {\em entanglement swapping} \cite{entang_swap}. This scheme is remarkable since we can entangle two qubits which never interacted in the past. The protocol for entangling the spin of two electrons is the following. We start with the initial state $\ket{\tilde \Psi^+}_{\sigma_1 k_1} \ket{\tilde \Psi^+}_{\sigma_2 k_2}$: for each electron the spin and orbital degrees of freedom are entangled as described previously. By performing a projective Bell measurement of the orbital qubits $\ket{k_1, k_2}$, the spin part of the two electron wavefunction becomes now entangled.
\putfig{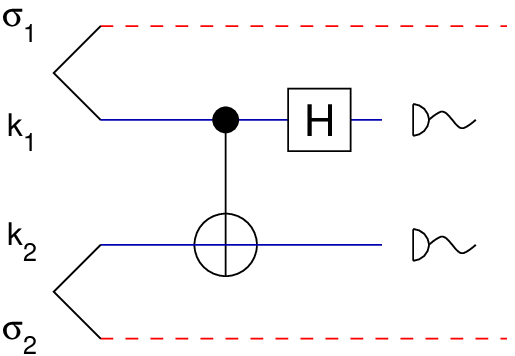}{4}
The joined $\sigma_i$ and $k_i$ lines in the left side denote that the corresponding qubits start in a maximally entangled state $\ket{\tilde \Psi^+}_{\sigma k}$. Since a projective measurement is involved, the protocol is probabilistic. The two spins will always end up in one of the four maximally entangled states, but the final state will depend on the outcome of the projective measurement of the mode qubits.

\noindent{\bf (b) Spin-orbital entanglement transfer.} Suppose we start with two electrons entangled only in modes in the following initial state $\ket{\up\up}_{\sigma_1 \sigma_2} \ket{00+11}_{k_1 k_2}$ (for simplicity we omit the normalization factor). By applying to this state the unitary $U=\SWAP(\sigma_1, k_1) \SWAP(\sigma_2, k_2)$, we transfer the entanglement from modes to spins, such that in the end the electrons are entangled in spins, but not in modes:
\be
\ket{\up\up}_{\sigma_1 \sigma_2} \ket{00+11}_{k_1 k_2} \stackrel{U}{\longrightarrow} \ket{\up\up+ \down\down}_{\sigma_1 \sigma_2} \ket{00}_{k_1 k_2}
\ee
Schematically, the quantum network is:
\putfig{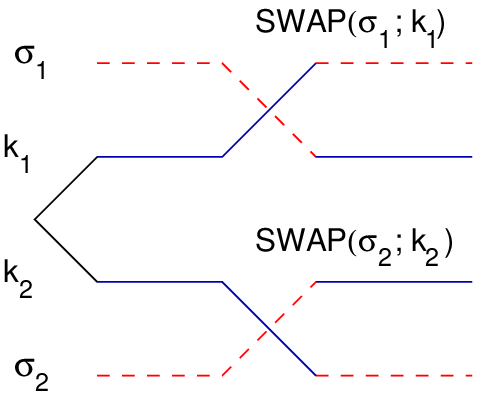}{4}
We stress that the $\SWAP(\sigma, k)$ gates are single particle gates, hence once the electrons have been entangled in modes, we can swap-entangle them in spins by performing only local operations on each electron.

Compared to the previous method, this scheme is deterministic as no measurement is required and it always outputs the same Bell state. Of course, the same protocol can be used in reverse for transferring the spin entanglement to mode entanglement, if the electrons start in a spin entangled state (e.g., a Cooper pair is injected in two quantum wires from a superconductor). Other protocols for entangle spins using mode degrees of freedom have been described in \cite{spin_space,bose_home}.

\subsection{Measurement}

Measurement is another important primitive in QIP, and as seen in the previous example, a Bell state measurement is used in the entanglement swapping protocol. We now discuss three measuring schemes.

\begin{figure}
\begin{tabular}{ll}
\cr
$\insertfig{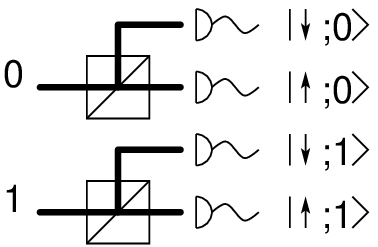}{3.5}$ \hspace{1cm} & $\insertfig{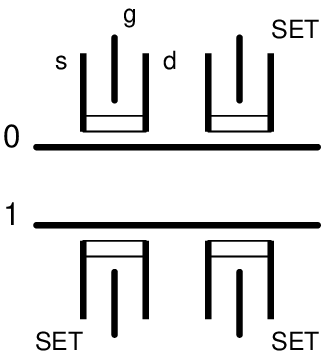}{3}$ \cr
\cr
(a) & (b) \cr
\end{tabular}
\caption{Two measurement setups: (a) Measuring both spin and mode qubits $\ket{\sigma; k}$. A PBS on each mode converts the spin into mode degrees of freedom such that the four basis states are separated into different channels. The detectors are single electron transistors (SETs) connected in anti-coincidence. (b) Measuring only the orbital degree of freedom $\ket{k}$. On each mode there are two charge detectors (SETs) mounted in coincidence, with a time window equal to the time-of-flight between the two; this avoids the random firing of one detector due to background noise. The top and bottom pair of SETs are connected in anti-coincidence.}
\label{measurement}
\end{figure}

Measuring the full state $\ket{\sigma; k}$ of a single electron can be performed with the scheme presented in Fig.~\ref{measurement}a. On each mode (0 and 1), a PBS converts the spin into orbital degrees of freedom thus separating the four possible states into distinct channels. Each channel is connected to a single electron transistor (SET) (wired in anti-coincidence since an electron can be in only one of the four states). This setup overcomes the difficult problem of single spin measurement by converting it into the easier problem of single charge measurement. A SET is composed of a single electron box (white rectangle in Fig.~\ref{measurement}b) situated between a source $s$ and a drain $d$. A plunger gate $g$ biases the single-electron island to work in the Coulomb blockade regime. The device is sensitive to single electron charges that propagate in the nearby quantum wire.

Measuring only the $\ket{k}$ state (the mode) can be done with the setup presented in Fig.~\ref{measurement}b. It is important to note that the measurement is non-absorbing, hence the electron still propagates in the quantum wires after the measurement. Since the detectors are not spin sensitive, a spin superposition state can survive the measurement. The collapsed state after the measurement is either $\ket{\sigma_0; 0}$ or $\ket{\sigma_1; 1}$; here $\ket{\sigma_i}= \alpha_i\ket{\up} + \beta_i\ket{\down}$ is still a spin superposition state.

In principle one can also measure only the spin state $\ket{\sigma}$, although it involves a more complicated setup. We first swap $\sigma$ and $k$, then we measure only $\ket{k}$, and finally we swap $\sigma$ and $k$ back.

\section{Applications}

\subsection{Orientable Stern-Gerlach} 

A mesoscopic polarizing beam-splitter is a versatile tool and can be used for both spin preparation and for {\em single} spin measurement. The original configuration \cite{pbs} is equivalent to a fixed Stern-Gerlach oriented along the $Oz$ axis. In practice however, we need to orient this device along an arbitrary axis.

Two important tasks require such flexibility. First, suppose we have a (stationary) source of spin current. We would like to test if the current is polarized or not, and if yes to determine the polarization direction ${\bf n}= (\cos \theta_0, \sin \theta_0)$. To do this, we need to be able to rotate the Stern-Gerlach in order to measure the spin along a variable direction. Let $p_\up(\theta)$ be the probability of measuring a spin up as a function of the angle $\theta$ of the Stern-Gerlach. From this function we can extract information about the polarization properties (e.g., the polarization degree) of the spin current. Thus, a completely unpolarized current has a constant $p_\up(\theta)=0.5$ for any angle, whereas a completely polarized current along direction $\theta_0$ will have $p_\up(\theta)= \cos^2(\theta-\theta_0)$.

A second task for which an orientable Stern-Gerlach is essential is testing Bell-CHSH inequality \cite{bell,chsh}. A pair of entangled spins is separated and at each end we measure the spin projection along two different directions. Here we need to measure the correlation function of the two spins $P(\vec n_1, \vec n_2)= \langle (\vec \sigma_1 \cdot \vec n_1) (\vec \sigma_2 \cdot \vec n_2) \rangle$ along the two directions $\vec n_1$ and $\vec n_2$; $\vec \sigma= (\sigma_x, \sigma_y, \sigma_z)$ is the vector of Pauli matrices and the subscript indicates the particle.

\begin{figure}
\putfig{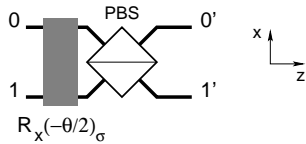}{4}
\caption{An orientable PBS. Measuring the spin along the direction $\vec n= (0, \sin\theta, \cos\theta)$ is equivalent to the above circuit. It first performs a spin rotation $\bbbone_k \otimes {\sf R_x}(-\theta/2)_\sigma$, then it measures the spin along the $Oz$ axis, i.e., in the computational basis, using the PBS. The Rashba gate (grey rectangle) is applied to both modes.}
\label{pbs_x}
\end{figure}

The solution is to rotate first the state and then to perform a spin measurement along the $Oz$ axis. Let $\ket{\psi'}= U\ket{\psi}$ be the rotated state. We determine the rotation $U$ by requiring $\bra{\psi} \sigma_{\vec n} \ket{\psi}= \bra{\psi'} \sigma_z \ket{\psi'}$, i.e., measuring the initial state $\ket{\psi}$ along the direction $\vec n= (0, \sin\theta, \cos\theta)$ is equivalent to measuring $\ket{\psi'}$ along $Oz$. A simple calculation gives $U= e^{-i\theta \sigma_x/2}$.

\subsection{Hyper-entangled electrons}

Hyper-entangled particles are quantum systems entangled in more than one degree of freedom and they have been experimentally realized with photons \cite{hyper}. Hyper-entanglement is a useful resource in several QIP protocols, like distillation and Bell state measurement \cite{kwiat}. Here we present a simple scheme for entangling two electrons in both spin and mode.

We start we two electrons entangled internally in spin-mode and we apply two swap operations $U'=\SWAP(\sigma_1, k_1) \SWAP(k_1,k_2)$ as in the following quantum network
\putfig{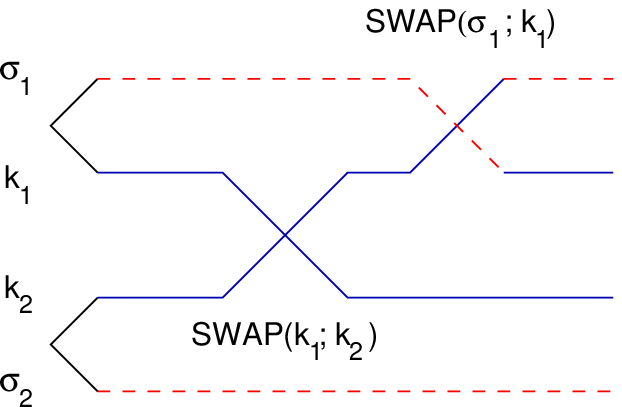}{5}
The mapping is:
\ba
\ket{\tilde \Psi^+}_{\sigma_1 k_1} \ket{\tilde \Psi^+}_{\sigma_2 k_2} &\stackrel{U'}{\longrightarrow}& 2^{-1} \ket{\up \up + i \down \down}_{\sigma_1 \sigma_2} \ket{00 + i 11}_{k_1 k_2} \cr
&=& \ket{\tilde \Psi^+}_{\sigma_1 \sigma_2} \ket{\tilde \Psi^+}_{k_1 k_2}
\ea
Thus, starting with two untangled electrons, we can transfer the internal entanglement (between spin and mode of the same particle) into hyper-entanglement. At the end of the operation, the two electrons are entangled both in mode and in spin. Note that the amount of entanglement is conserved: the initial 2 ebits (entanglement bits) are redistributed into spin and mode entanglement.

Experimentally, the major difficulty is the first $\SWAP(k_1,k_2)$ gate since it involves both electrons, and hence synchronization. The initial entangled state and the final $\SWAP(\sigma_1,k_1)$ involve only single particle gates and are easier.

\section{Conclusions}

In this article we presented a complementary, top-bottom approach to the usual view of spintronic devices. Rather than focusing on the detailed microscopic interactions, we assumed a small set of building blocks (like Rashba-active regions, beam-splitters, Aharonov-Bohm etc) from which we can obtain universality. That is to say, using these blocks we can construct any unitary transformation on a given qubit system.

One of the motivations of our approach was to find a solution to a reverse engineering problem. We start with a ``black-box'' with some desired properties (e.g., spin-spin entangler, polarizing beam-splitter etc). Assuming we have a finite set of resources (beam-splitters, Aharonov-Bohm phases, Rashba-active regions etc) the problem is to find a blueprint of the desired apparatus in terms of the set of available interactions. The solution we proposed here is to think the device as a network of quantum gates acting on different degrees of freedom (spin and orbital) and to find the simplest decomposition in terms of the available resources.

Using this framework, we have shown how to design several devices, including spin-spin entanglers and an orientable mesoscopic PBS, which can act as a flexible Stern-Gerlach apparatus for spin preparation and measurement.

A possible extension of the present scheme is to include interband coupling. In this case, there will be an extra qubit (or qudit, in general) which takes into account the two (or more) bands present in each wire (mode). A single electron will then encode three qubits, one for each degree of freedom: spin, orbital and band. The interactions will be more complex and the set of elementary gates will be enlarged by interband coupling interactions, among others.

We hope that the approach presented here will encourage the design of several spintronic devices by providing a unified architecture in terms of quantum gates.

\noindent {\bf Acknowledgements.} I am grateful to Anca Popescu for carefully reading the manuscript and for valuable comments.

\end{document}